\documentclass[aps,amsmath,amssymb,superscriptaddress,nofootinbib]{revtex4}

\usepackage{graphicx}
\usepackage{hyperref}
\usepackage{epsf}

\newcommand{\ba}{\begin{align}}
\newcommand{\ea}{\end{align}}
\newcommand{\la}{\label}
\newcommand{\be}{\begin{equation}}
\newcommand{\ee}{\end{equation}}
\def\12{\frac{1}{2}}
\newcommand{\p}{\partial}

\begin{document}

\title{Nonlinear Dynamics of Quantum Systems and  Soliton Theory}
\author{E. Bettelheim}
\affiliation{James Frank Institute, University of Chicago, 5640 S.
Ellis Ave. Chicago IL 60637.}
\author{A. G. Abanov}
\affiliation{Department of Physics and Astronomy,
Stony Brook University,  Stony Brook, NY 11794-3800.}
\author{P. Wiegmann}
\affiliation{James Frank Institute, University of Chicago, 5640 S.
Ellis Ave. Chicago IL 60637.}
\affiliation{Also at Landau Institute of Theoretical Physics.}

\begin{abstract}
We show that space-time evolution of  one-dimensional fermionic
systems is described by nonlinear equations of soliton theory. We
identify a space-time dependence of a matrix element of fermionic
systems related to the {\it Orthogonality Catastrophe}  or
{boundary states} with the $\tau$-function  of the modified
KP-hierarchy. The established relation allows to apply the
apparatus of soliton theory to the study of non-linear aspects of
quantum dynamics. We also describe  a {\it bosonization in
momentum space} - a representation of  a fermion operator by a
Bose field in the presence of a  boundary state.
\end{abstract}
\date{\today}

\maketitle


\section{Introduction}

In the seminal paper by Date,  Jimbo,  Kashiwara and Miwa \cite{1}
Sato's approach \cite{Sato1} to integrable hierarchies of soliton
equations has been formulated in terms of auxiliary fermions. In
this formulation the hidden symmetries of soliton equations become
explicit symmetries of free fermions, the modes of which are
labelled by one-dimensional parameter with respect to the algebra
of fermionic bilinear forms -- $gl(\infty)$. It was shown that the
$\tau$-function of soliton theory can be expressed as a certain
matrix element of 1D-fermions. The paper \cite{1} and subsequent
works of the Kyoto school (see \cite{2} and references therein)
laid out a mathematical foundation of  {\it bosonization} --- a
commonly used representation of fermionic modes in terms of a Bose
field. These works established a deep relation between quantum
physics and classical nonlinear equations.

In this paper we adopt this theory to a particularly interesting
area of electronic physics - dynamics of fermionic coherent states
propagating in real space and time generated by the Galilean
Hamiltonian
\begin{equation}\label{H}
H=\sum_p \frac{p^2}{2M}\psi^\dag_p\psi_p.
\end{equation}
We will show that the real time evolution of important objects of
electronic physics  also obeys classical non-linear soliton
equations. Therefore the powerful apparatus of soliton theory can
be used to study complex aspects of non-equilibrium  physics of
effectively one-dimensional quantum systems.  We will report some
applications of the results of this paper to non-linear electronic
transport elsewhere. Here we focus on the formal connection of
soliton theory and quantum dynamics. In this course we encounter
an alternative {\it bosonization} with respect to the boundary
state, but now in momentum space.

Before formulating our main results, we introduce notations and
give a brief review of some important facts on a connection
between solutions of soliton equations and fermionic matrix
elements. The literature on the subject is enormous, we restrict
the reference list to the original papers \cite{1,2}.

\subsection{Bilinear identity}

Consider a Fermi-system on a unit circle or, equivalently, 1D
fermions subject to periodic boundary conditions. We let $\psi_p$
be a fermionic mode of the fermionic field $\psi(x)=\sum_p e^{ipx}
\psi_p$, labelled by integer momenta, $p$. The conjugated field is
given by $\psi^{\dagger}(x)=\sum_p e^{-ipx} \psi^{\dagger}_p$ with
basic (anti-) commutation relations $ \left \{ \psi_{p} , \psi_{q}
\right\} = \left\{\psi^{\dagger}_{p}, \psi^{ \dagger}_{ q }
\right\}=0$ and $\left \{ \psi_{p},\psi^{ \dagger}_{q} \right \} =
\delta_{p,q}$. The ground state of the system  is a filled
Fermi-sea such that all modes in the interval between the
Fermi-points are filled. We will be interested only in one, say,
the right chiral sector where all essential modes are concentrated
around the Fermi point $+p_F$. It is convenient to count modes
from this Fermi point and change $p\to p-p_F$. Then the  vacuum is
the state with all non-positive momenta filled by fermions. It is
defined as the state $|0\rangle$ such that:
\begin{equation}\label{vac}
\psi_p|0\rangle = 0,\;\; {\rm if}\; p> 0,\quad \quad \psi_p^\dag|0
\rangle=0 , \; \;  {\rm if}\; p\leq 0.
\end{equation}
A state, where in addition to the filled negative modes, $m>0$
($-m>0$) consecutive positive (negative) modes are also filled
(empty) is denoted by $|m\rangle $ ($| - m\rangle $).

Excited states are obtained by a repeated action of fermionic
bilinear operators $\sum A_{pq}\psi^\dag_p \psi_q$ creating
particle-hole excitations on the  vacuum. These operators  form an
infinite dimensional algebra $gl(\infty)$. Coherent states of this
algebra are the states obtained by the action of a group element
on the vacuum
\begin{equation}\label{g}
\langle m|g=\langle m|e^{\sum_{p,q} A_{pq}:\psi^\dag_p \psi_q:} .
\end{equation}
A  coherent state represents  {\it particle-hole wave-packets}. It is characterized
 by a infinite matrix  $A_{pq}$. The colon in (\ref{g}) and throughout the paper
denotes the normal ordering with respect to the vacuum (\ref{vac}),
$:\psi^\dag_p \psi_q: \;= \psi^\dag_p \psi_q-\langle 0|\psi^\dag_p \psi_q |0\rangle$.

Within the algebra of fermionic bilinears one distinguishes two
commutative subalgebras. One  is generated  by positive $k>0$
(or by negative $k<0$) modes of the current
\begin{equation} \label{J}
J_k=\sum_p :\psi^\dag_p \psi_{p+k}:,\quad [J_k,\, J_{-k'}] = k
\delta_{kk'}.
\end{equation}
This subalgebra is central in the construction of Ref. \cite{1,2}.
The other commutative subalgebra will be  central in the
construction of the current paper. It is defined in (\ref{HK}).

An important result \cite{1,2} is that the matrix element
\begin{align} \label{tau1}
    \tau_m({\bf t})=\langle m| g\; e^{- \sum_{k>0} t_k J_{-k}} |m\rangle
\end{align}
considered as a function of an infinite number of continuous
parameters ${\bf t}=\{t_k\}$ and of an integer parameter $m$ is
the tau-function of an integrable hierarchy.  $\tau_m$ depends
also on the state $\langle g|$ (or on the matrix $A_{pq}$), but
this dependence will be omitted in notations throughout the paper.
This means that the family of matrix elements $\tau_m({\bf t})$
obeys a bilinear identity:
\begin{equation} \label{Hirota}
    \oint_0 e^{\sum_{k>0}(t_k-t'_k)z^k} \tau_{m} \left( {\bf t} - [z]
    \right)   \tau_{m + n} \left( {\bf t'} + [z] \right) z^n\, dz=0,
\end{equation}
where  ${\bf t}\pm  [z] \equiv  \left\{ t_k\pm z^{-k}/k
\right\}\;k=1,2,\dots$ and ${\bf t}$, ${\bf t'}$ are two
independent sets of parameters. The integral goes along a small
contour around  the origin.

The bilinear identity (\ref{Hirota}) encodes an infinite set of
hierarchically structured non-linear differential equations in the
variables (flows) $t_k$. For a given integer $n$, this set is
called the $n$-MKP (modified KP) hierarchy. The differential
equations with respect to the flows $t_k$ in a bilinear (Hirota)
form are obtained by expanding the integrand of (\ref{Hirota}) in
powers of $z^{-1}$ and setting $t'_k=t_k$.

The KP-hierarchy ($n=0$) consists of  a set  of differential
equations on $\tau_{m}$. The first equation of the KP-hierarchy
reads
\begin{align}\label{KP}
(D_1^4+3 D_2&^2-4D_1 D_3)\tau_m\cdot\tau_m=0,
\end{align}
where $D_k$ is the Hirota derivative in the variable $t_k$ defined
as $D_k^n\,a\cdot b=(\p_{t_k}-\p_{t'_k})^na({\bf t})b({\bf
t'})|_{t_k'=t_k}$. This equation is a bilinear form of the
Kadomtsev-Petviashvili equation of plasma physics \cite{1}. It
gave the name (KP) to the hierarchy.

The MKP-hierarchy  involves a discrete evolution in $m$.  The
first equation of the 1-MKP-hierarchy reads
\begin{align} \label{MKP}
(D_2+D_1^2)\tau_{m+1}\cdot\tau_m=0.
\end{align}

\subsection{Evolution in the space of parameters}

Some physical applications arise if  $t_k=-t_{-k}^*$. We write
$t_{k} = i\delta t A_{-k}$ and notice that the operator $\sum_k
t_k J_k=i\delta t\int \, A(x)\rho(x)(dx/2\pi)$  represents the
action of  the scalar electromagnetic potential $A(x)=i\sum_k
t_{-k} e^{ikx}$
\begin{align}
e^{ - \sum_k t_k J_k} = e^{-i\delta t\int A(x) \rho(x) \,
dx},\quad \rho(x)=:\psi^\dag (x)\psi(x):,
\end{align}
which was turned on instantaneously for a short period $\delta t$.
This matrix element is  relevant, e.g., for the process when  an
electronic system being initially in an excited state $\langle g|
= \langle 0|g$ is instantaneously hit by  radiation. The
tau-function then is the amplitude of the probability to find  a
system  in the vacuum (ground) state.

An alternative interpretation of the formal expression
(\ref{tau1}), common  in the  literature, emphasizes the
commutativity of flows with respect to parameters $t_k$. In this
interpretation the operator $e^{\sum_k t_k J_{-k}}$ is viewed as
an evolution operator with a set of ``times'' $t_k$ corresponding
to a family of mutually commuting ``Hamiltonians'' $J_k$. The
excited state $\langle g|=\langle m|g$ evolves in ``times''. The
tau-function is an overlap between the ``evolved'' excited state
$\langle g({\bf t})|$ and the vacuum
\begin{equation} \la{tau11}
\tau_m = \langle g({\bf t}) | m\rangle, \quad \langle g({\bf t})|
=\langle m|g\; e^{-\sum_k t_k J_{-k}}.
\end{equation}
This terminology, although common,  is somewhat confusing.  Of
course, the ``Hamiltonians'',  $J_{k}$ are not Hermitian and there
is no energy associated with this evolution. The ``evolution''
occurs in the space of parameters. This may be a reason why
applications of the powerful mathematical apparatus of soliton
theory to electronic physics have been limited.

In the next subsection we introduce the tau-function as a matrix
element of the real-time evolution generated by the   Hermitian
Hamiltonian (\ref{H}).

\subsection{Evolution in real time}

In physics one is interested in the real-time evolution driven by
a Hermitian positive Hamiltonian as (\ref{H}) rather than currents
(\ref{J}).

Can the theory of solitons be used to study the dynamics of
fermions in the real space-time, when the  evolution is described
by a unitary operator
\begin{equation} \label{HP}
e^{-iHt+iPx}
\end{equation}
corresponding to a positive energy  (\ref{H}) and momentum
\begin{equation} \label{P}
 P=\sum_{p \in I} p :\psi^\dag_p \psi_p:.
\end{equation}
In this case the vacuum is selected as a ground state of the
Hamiltonian $$H|0\rangle=0.$$ Indeed,  the set of Hermitian
Hamiltonians
\begin{equation} \la{HK}
H_k=\sum_{p \in  I} p^k :\psi^\dag_p \psi_p:,\quad k=1,2,\dots
\end{equation}
generates another commutative subalgebra of fermionic bilinears.
Here we denote  the set  of quantized momenta for the system by
$I$.

Surprisingly, this question has never been addressed
systematically. The most relevant papers we know are Refs.
\cite{SIK,SMJ} where the relation of time dependent Green
functions of impenetrable bosons and Green functions of Ising
model to the Painlev\'{e} equations has been established.

In this paper  we will show that, indeed, an action of the
commutative Hamiltonians  $(\ref{HK})$ generates an integrable
hierarchy. Specifically, we show that the matrix element
\begin{equation}  \label{tau2}
    {\tau}_m({\bf t})=\langle 0| g \; e^{- \sum_{k>0}  t_k H_k}
    :e^{-(a+m)\,\varphi(x)}:|0\rangle,\quad 0<|a|<1/2, \quad
    m\;\rm{is\; integer}
\end{equation}
obeys the MKP hierarchy \footnote{Here we take $a\neq 0$. Although
the tau-function  obeys the same $a$-independent equations, it
degenerates at $a=0$ and can be computed by elementary means.}.

Namely, the tau-functions $\tau_m({\bf t})$, similarly to
(\ref{tau1}), satisfy a set of equations analogous to (\ref{KP},
\ref{MKP}). These equations are generated by the identity:
\begin{equation} \label{ContinuousHirotaH}
\oint_0 e^{\sum_{k>0}(t_k - t'_k)z^k }
\frac{\Gamma(z+m+n)}{\Gamma(z+m)} \tau_{m} \left( {\bf t} - [z]
\right) \tau_{m+n} \left( {\bf t'} + [z] \right)  dz=0,
\end{equation}
which is a modified form of (\ref{Hirota}). The factor
$\frac{\Gamma(z+m+n)}{\Gamma(z+m)}= \prod_{j =m}^{m + n -1} ( z +
j ) $ replaces $z^{n}$ in (\ref{Hirota}) \footnote{ The difference
between (\ref{ContinuousHirotaH}) and (\ref{Hirota}) is
inessential insofar as the former may be transformed into the
latter by the transformation $\tau_m \to \tau_m e^{\sum_{k\geq0}
V(m+k)}$, where $V(z)=\sum_k t_k z^k $, however, this
transformation changes analytical properties of the tau-function
as a function of the ``times".}.

In the eq. (\ref{tau2}) the operator $\varphi$ is defined as a
Bose field
\begin{align} \label{phi}
    \varphi(x)=\sum_{k \neq 0} \frac{1}{k}e^{ikx}J_k +iJ_{0}x
\end{align}
and
\begin{equation}
    :e^{ a\varphi(x)}:=e^{a \varphi_-(x)} e^{iaxJ_0} e^{a \varphi_+
    (x)},
\end{equation}
where $\varphi_\pm (x)= \sum_{\pm k >0}(e^{ikx}/k) J_k $ contain
only positive or negative modes respectively. Note that the flow
$t_1$ just shifts the coordinate $x$ by $it_1$ as $e^{t_{1}H_{1}}
\varphi(x) e^{-t_{1} H_{1}} = \varphi ( x + it_{1} )$, so we may
set $x=0$ in some formulas below.  In physical applications we
identify
\begin{equation} \label{t}
    {\rm time}: \; t = it_2, \quad \mbox{space coordinate in the
    Galilean frame}: x-v_F t = - it_1,
\end{equation}
where $v_F=p_F/M$ is a Fermi velocity.

The operator $:e^{ a\,\varphi(x)}:$ in (\ref{tau2}) is called
{\it a boundary condition changing operator}. This operator
acting on a vacuum creates {\it a boundary state}
\begin{equation}
    |B_m(x)\rangle=:e^{-(a+m) \varphi(x)}:|0\rangle.
\end{equation}
The meaning of the boundary condition changing operator is
clarified by the action of the counting operator
 $n(y)=\int_0^y(:\psi^{\dag}(y')\psi(y'):-(a+m))dy'$
 on the boundary state
\begin{equation}
    \langle B_{m}(x)|n(y)|B_m(x)\rangle
= (a+m)\quad {\rm  if}\; 0<x<y,\quad{\rm or}\;\; 0\quad {\rm otherwise}.
\end{equation}
In other words, the boundary condition changing operator moves
particles towards the point $x$ creating an increase of the
density $\rho(y)=\partial_{y}n(y)$ at that point.

Boundary states  play an important role in 2D-critical phenomena
in systems with boundaries.  The connection of the latter to
1D-fermions is following. The fields (e.g.  $\psi(x)$ and
$\varphi(x)$) are originally defined on the unit circle. One views
the circle as a contour embedded in the complex plane $z$. The
fields and their  differentials can be analytically extended to
the exterior of the circle and to its interior. Instead of
considering fields defined at the point $z'$ in the interior of
the circle one views them as defined at the point $z= 1/\bar z'$
of the circle's exterior (Schwarz reflection). From this point of
view the circle $z=e^{-ix}$, $x \in \mathbb R$,  is seen as a
boundary of the holomorphic 2D-field theory defined in its
exterior.


In electronic physics boundary states appear in numerous problems
related to the Fermi-edge singularities, quantum  impurity
problems  and other phenomena related to Orthogonality Catastrophe
(\cite{OC}, see, e.g., \cite{AA}). We will discuss some
applications to electronic physics elsewhere \cite{BAW}.

In terms of the boundary state, our tau-function (\ref{tau2}) reads
\begin{equation} \la{tau22}
\tau_m({\bf t}) = \langle {G\bf(t)} | B_m \rangle, \quad
\langle{G\bf( t})| = \langle 0|g\; e^{ -\sum_{k>0}t_k H_k}.
\end{equation}

The tau-function depends on two parameters: a fraction $a$,   and
an integer $m$. Below we fix $a$  and consider $m$ as a running
parameter.

One can  think of this matrix element as follows. An excited
coherent state $\langle g|$ evolves up to a time $t$ and then is
measured at a point $x$ by a projection onto a boundary state
$|B_a(x)\rangle$.

A comment is in order. The presence of the boundary  operator $:e^{a\,
\varphi(x)}:$ changes the momentum quantization. The eigenvalues
$p$ of the momentum operator (\ref{P}) appearing in (\ref{P},
\ref{HK}) are no longer integer, but are integers shifted
by $a$:
\begin{equation} \label{19}
p\in {\mathbb Z}+a,
\end{equation}
that is $I$ in (\ref{P}) and (\ref{HK}) is given by $I = {\mathbb
Z} + a$.

An obvious generalization of this result is a passage from the KP
to the 2D Toda hierarchy \cite{Ueno-Takasaki}. The matrix element
$$\langle B_m|e^{-\sum_k\bar t_k H_k}g e^{-\sum_k t_k H_k}|B_m\rangle$$
obeys the 2D-Toda hierarchy with respect to two sets of parameters
$t_k$ and $\bar t_k$ and a discrete parameter $m$.  Other obvious generalization occurs
when fermions are placed on a spatial lattice  so that the energy
spectrum is a periodic function of momentum. We  do not discuss
these generalizations here.

The modified bilinear identity (\ref{ContinuousHirotaH}) generates
an infinite set of Hirota bilinear equations. It, therefore, obeys
KP hierarchy with respect to infinitely many flow parameters
$t_{k}$. It also obeys the MKP hierarchy with respect to a
discrete parameter $m$ characterizing the boundary state. In a
special case when only two physical flow parameters  (\ref{t})
$t_2$ and $t_1$ are present we obtain the 1-MKP equation. In the
Galilean frame it reads
\begin{equation} \label{MKP1}
(iD_t + D_x^2 - 2imD_x)\tau_{m+1}\cdot\tau_m=0.
\end{equation}
The independent variables $x$ and $t$ in this  equation are
physical space and time coordinates of a 1D fermionic system.
Among the hierarchy, this equation seems to be the most important
for the study of the dynamics of electronic systems. The two
functions $\tau_m$ and $\tau_{m+1}$ are not independent but are
connected by certain analytical properties, thus making equation
(\ref{MKP1}) closed. In the most interesting situations
(corresponding to particular choices of coherent state $g$) the
analytical conditions are explicit and correspond to a certain
reduction of the MKP-hierarchy. We consider coherent states having
particle-hole symmetry and the corresponding reduction in Appendix
\ref{app:reduction}.

We note here that the matrix element (\ref{tau2}) give solutions
of the MKP hierarchy, which are essentially different from the
ones given by (\ref{tau1}). For example, we show below that
(\ref{tau2}) has only positive Fourier modes 
with respect to all odd flows $it_1,it_{3},\ldots$. This analytical
condition is restrictive. In particular, it does not admit soliton
solutions. In contrast, it is well known that (\ref{tau1}) can
generate soliton solutions (see Sec.\ref{multiphase}).

Some mathematical constructions related of the matrix element
(\ref{tau2}), although devoted to different problems, have
appeared recently in papers by A. Orlov \cite{orlov-2003-} and by
A. Okounkov, R. Pandharipande, and N. A. Nekrasov
\cite{okounkov-2000-}.

\section{Hirota Equation and Bosonization}

Before we proceed,  we present another form of  the bilinear
identity  (\ref{ContinuousHirotaH}) for a  particular case $n=1$.
Setting ${\bf t} = {\bf t^{(0)}} + [p]$, ${\bf t'} = {\bf t^{(0)}}
- [q]$ the factor $(z+m)e^{\sum_k (t_k-t'_k) z^k}$ becomes $
\frac{(z+m)p q}{(q-z ) (p-z)}$.  The integrand has three poles at
$z=0,p,\,q,\infty$ and can be computed. It gives the discrete
Hirota equation
\begin{equation}\label{DH}
    (p+m)\, \tau_m\cdot \tau_{m+1}([p]-[q]) - (q+m) \,
    \tau_m([p]-[q])\cdot \tau_{m+1} - (p - q)\, \tau_m ([p])
    \cdot\tau_{m+1} (-[q]) =0,
\end{equation}
where we have dropped the term $\bf  t^{(0)}$ in each of the
arguments of the $\tau$ functions. We stress that $p$ and $q$ are
shifted integers $p,q\in {\mathbb Z}+a$.

Hirota's equation, written in the discrete  form (\ref{DH})
appears as a particular specification of the integral form of the
bilinear identity (\ref{Hirota}). In fact, they carry the same
information as an integral form of the bilinear identity written
for an arbitrary $n$. The latter can be obtained from the former
\cite{Sato1}. In the rest of this section we will derive the
discrete form of Hirota's equation (\ref{DH}) for tau-functions
defined in (\ref{tau2}). It will be equivalent to proving
(\ref{ContinuousHirotaH}).

Our main equation (\ref{MKP1}) directly follows from Hirota
equation written in the form (\ref{DH}). It appears in the leading
order expansion at large $q$ and $p$, and setting $q=-p$.

\subsection{Bosonization in  coordinate space and the tau-function
$\tau_m=<g({\bf t})|m>$ }

First, we  show how to derive the discrete Hirota equation
(\ref{DHJ}) for the the conventional matrix element matrix
(\ref{tau1}). Then we derive the discrete Hirota equation
(\ref{DH}) for our  tau-function (\ref{tau2}). The first
derivation is well known  \cite{Hirota:PRL,1,Sato1,2}. We,
nevertheless, present it here in order to use it as a guide for
the following derivation of Hirota's equation for the real time
evolution. The comparison of these two derivations is instructive.

Applying the same trick as  is the previous section  to
(\ref{Hirota})  we obtain a discrete Hirota equation for the
tau-function (\ref{tau1})
\begin{align}\label{DHJ}
    p\, \tau_m\cdot \tau_{m+1}([p]-[q]) - q \, \tau_m([p]-[q]) \cdot
    \tau_{m+1} - (p-q)\, \tau_m ([p]) \cdot\tau_{m+1} (-[q]) =0.
\end{align}

To derive it one needs two technical ingredients:
\textit{bosonization} and the \textit{Wick's theorem}.

\subsubsection*{Bosonization formulas}
The Bosonization formulas express the action of fermion operators in
 terms of bosons. We will need the following formulas (briefly derived  below)
\begin{equation}
Ê\la{BF1}
ÊÊÊ \psi^\dag(x)\psi(y)|m\rangle = \frac{ 1}{ 1 - e^{i (x-y) } }
ÊÊÊ :e^{ \varphi (y) - \varphi(x)}:|m\rangle,
\end{equation}
\begin{equation}
Ê\label{bf1}
ÊÊÊ \psi(x) |m+1 \rangle = e^{ix} :e^{Ê \varphi(x)}: | m\rangle,
ÊÊÊ \quad \quad \psi^\dag(x) | m \rangle =
ÊÊÊ :e^{-\varphi(x) }: | m + 1 \rangle .
\end{equation}

Let us now use these bosonization formulas to ``fermionize''Ê the
tau-function $\tau_{m+1}([p] - [q])$. We set $q=e^{ix}$ and
$p=e^{iy}$, and write the tau-function $\tau_{m+1}([p] - [q])=
\langle m+1 | g\; e^{-\sum_{k}t_k^{(0)} J_{-k} } :e^{
\varphi(y) -\varphi(x)}: | m+1\rangle$, appearing in
(\ref{DHJ}), in terms of fermion operators:
\begin{equation}
ÊÊÊ \tau_{m+1}([p]-[q]) =e^{i(m+1)(x-y)}(1 - e^{i(x-y)})
ÊÊÊ \langle m+1|Ê g({\bf t}) \psi^\dagger(x) \psi(y)
ÊÊÊÊ |m+1\rangle, \quad g({\bf t})=g\; e^{-\sum_{k} t_k^{(0)}J_{-k} }.
\end{equation}

\subsubsection*{Wick's theorem}

We apply Wick's theorem to four fermion matrix elements. Let
$\chi_i$ be a linear combination of fermionic modes $\chi_i =
\sum_q v^{(i)}_q \psi_q$,   and $|G_2\rangle$  and $\langle  G_1|$
are coherent states (i.e. states  of the form (\ref{g})). The
general form  of  Wick's theorem for four fermion matrix elements
reads
\begin{equation}\label{wickschi}
    \langle G_1|\chi^{\dag}_{1}\chi_{2}\chi^{\dag}_{3} \chi_{4} | G_2
    \rangle \langle G_1|G_2\rangle = \langle G_1 | \chi^{\dag}_{1}
    \chi_{2}| G_2 \rangle \langle G_1 | \chi^{\dag}_{3} \chi_{4} | G_2
    \rangle + \langle G_1 | \chi^{\dag}_{1} \chi_{4}| G_2 \rangle
    \langle G_1| \chi_{2}\chi^{\dag}_{3}|G_2\rangle.
\end{equation}
\subsubsection*{A proof of the Hirota bilinear identity (\ref{DHJ})}
We  apply Wick's theorem  to the matrix element:
\begin{align}
    & < m| \psi_{ m+1 }\, g({\bf t})\, \psi^\dagger(x) \psi(y)
    \psi^\dag_{ m+1} | m\rangle  \langle m| g({\bf t}) | m\rangle
    = \langle m+1| g({\bf t})|m+1\rangle \langle m| g({\bf t})\,
    \psi^\dagger(x) \psi(y)| m\rangle
 \nonumber \\
    & + \langle m+1|  g({\bf t})\, \psi^\dagger(x)
    |m\rangle \langle m| g({\bf t})\, \psi(y)  | m+1\rangle.
 \la{wick1}
\end{align}
Eq.  (\ref{DHJ}) emerges if one  writes  each of the matrix
elements appearing in this identity as $\tau$-functions with
shifted arguments, by further use of the bosonization formulas
(\ref{BF1},\ref{bf1}) \footnote{This proof appeals to a general
view on tau functions as Pl\"ucker coordinates of the Grassmann
manifold \cite{Sato1}.} .

\subsubsection*{Derivation of bosonization formulas (\ref{BF1},\ref{bf1})}
It will be instructive to have a short proof of bosonization
formulas (\ref{BF1},\ref{bf1}) as we are going to use a similar
approach  in the next section. In the following we set for
simplicity $m=0$.

Let us project  the r.h.s.  and the l.h.s.  of (\ref{BF1}) onto a
general state with a given number of particles
$\langle\{x_i,\,y_i\}|=\langle 0| \left(\prod_{i=1}^{N}
{\psi}^\dagger(x_i ) {\psi}(y_i)\right)$, consisting of $N$
particle and $N$ hole excitations with coordinates $x_i,\,y_i$,
and compare the results.

First of all, let us use Wick's theorem to calculate the overlap
between the state with $N$ particle-hole pairs  and the vacuum.
This overlap  is a determinant of a single-particle Green's
function $\langle 0|\psi^{\dag}(x)\psi(y)|0\rangle =(1-e^{i(x-y)})^{-1}$
\begin{equation}  \label{Cauchy1}
    \langle\{x_i,\,y_i\}| 0\rangle
    = e^{i \sum_{i=1}^{N} y_i} \det_{i,j \leq N}
    \left(\frac{1}{e^{i y_i}-e^{i x_j}} \right)
    =e^{i \sum_{i=1}^{N} y_i}
    \frac{\prod_{i< j \leq N} (e^{i y_i}-e^{i y_j}) (e^{i x_j}-e^{i x_i}) }
    {\prod_{i,j\leq N} (e^{i y_i}-e^{i x_j})},
\end{equation}
where  we used the multiplicative form of the Cauchy determinant
\begin{equation} \label{C}
    \det_{i,j \leq N} \left(\frac{1}{q_i-p_j}\right)
    =\frac{\prod_{i< j \leq N} (q_i-q_j) (p_j-p_i) }
    {\prod_{i,j\leq N} (q_i-p_j)}.
\end{equation}

The l.h.s. of (\ref{BF1}) is computed by extending the previous formula to $N+1$ particle-hole pairs and setting
$x_{N+1}=x$ and $y_{N+1}=y$
 \begin{eqnarray}
    \langle\{x_i,\,y_i\}| {\psi}^\dagger( x) {\psi}(y)  |0\rangle=
 \frac{e^{iy}}{e^{iy}-e^{ix}}
    \prod_{i\leq N} \frac{(e^{iy}-e^{iy_i})(e^{ix}-e^{ix_i})}
    {(e^{iy}-e^{ix_i})(e^{ix}-e^{iy_i})} \;
     \langle \{x_i,\,y_i\}|0\rangle .
  \label{Cauchy}
\end{eqnarray}

The r.h.s.  of (\ref{BF1}) can be computed by  (i) noticing that
$e^{b\varphi}|0\rangle=e^{b\varphi_-}|0\rangle$,   (ii) pulling
the vertex operator $e^{b\varphi_{-}}$ to the left  using the
identity
\begin{align}
 \la{28}
    e^{-b \varphi_{-}(x)} \psi^\dagger(x_i)\psi(y_i) e^{b\varphi_{-} (x)}
    =\left( \frac{ e^{ix} -e^{iy_i}}{e^{ix} - e^{ix_i} }\right)^{b}
    \psi^\dagger(x_i)\psi(y_i)
\end{align}
and then (iii) using $\langle 0|e^{b\varphi_{-} (x)}=0$. As a result we get
\begin{align} \label{bosonizationproof1}
&\langle \{x_i,\,y_i\}| :e^{\varphi (y)-\varphi(x)}: |0\rangle =
\prod_{i\leq N} \frac{(e^{iy}-e^{iy_i})(e^{ix}-e^{ix_i})}
{(e^{iy}-e^{ix_i})(e^{ix}-e^{iy_i})} \; \langle
\{x_i,\,y_i\}|0\rangle.
\end{align}

Comparing eqs. (\ref{bosonizationproof1}) and  (\ref{Cauchy})
gives  the bosonization formula (\ref{BF1}). Eq. (\ref{bf1}) can
be derived in a similar way, or by  setting $x=-i\infty$ in
(\ref{BF1}) and using the fact that
$\psi^\dag(-i\infty)|m\rangle=|m+1\rangle$. In all formulas $x$
and $y$ carry small positive imaginary parts.

\subsection{Bosonization in  momentum  space and the tau-function
 $\tau_m=< {G\bf(t)}| B_m>$ }

Now we show  that the matrix element  (\ref{tau2}) obeys the
Hirota equation (\ref{DH}). The steps of the derivation in this
section follow the  steps  of the previous section, but
technically different.  We repeat that in this section all momenta
are shifted integers  $p,\,q\in {\mathbb Z}+a$.

The important property of the boundary state is displayed  by the
form of its wave-function in momentum space. We have $\langle
0|B_{m}\rangle=1$ and two important formulas
\begin{align}
 \label{Ga}
    &  \langle 0| \psi^{ \dagger  }_q  \psi_p  |B_{m}\rangle
    =f^{+}_m(p)\frac{1}{p-q} f^{-}_m(q),\quad q\leq a < p,
 \\
    & \frac{p+m}{q+m}\langle 0| \psi^{ \dagger  }_q \psi_p |B_m\rangle
    = \langle 0| \psi^{\dagger  }_q \psi_p |B_{m+1}\rangle,
 \label{29}
\end{align}
where
\begin{align} \label{efs}
& f^{+}_m(p)  = \frac{ \Gamma(p+m)}{\Gamma (m+a)\Gamma(p-a)}, \nonumber\\
& f^{-}_m(q) =\frac{\Gamma(1-q-m)}{\Gamma (1-m-a)\Gamma(1-q+a)}.
\end{align}
The matrix element (\ref{Ga}) is also non zero at  $p=q \leq a$,
where it is equal to $1$\footnote{The formula (\ref{Ga}) is
understood as being  analytically continued from the real axis to
the complex plane of  $p$ and $q$. In particular, at $p=q$ it
should be understood as a limit $p\to q$ at fixed $q$. }. Eq.
(\ref{29}) follows from (\ref{Ga},\ref{efs}) with  the help of the
formula
\begin{equation} \label{f}
f^{+}_m(p)f^{-}_m(q) \, \frac{ p + m}{q + m} = f^{+}_{ m + 1 }(p)
f^{-}_{ m + 1 }(q).
\end{equation}
We prove these formulas at the end of this section.

The formula (\ref{Ga}) is  related to a phenomenon referred to as
``Orthogonality Catastrophe'' \cite{OC}. Namely, if the length of
the ring is large, i.e, in the limit $p, -q\gg |m|$ the formula
(\ref{Ga}) reads
\begin{align} \label{Ga11}
\langle 0| \psi^{ \dag}_q \psi_p  |B_m\rangle =
(-1)^m\frac{\sin(\pi a) }{\pi} \frac{1}{p-q} \left( \frac{p}{-q}
\right)^{a+m}.
\end{align}
The overlap between the excited state and the boundary state
vanishes as the momentum of the excited particle approaches the
Fermi-point $\langle 0| \psi^{\dag}_q  \psi_p  |B_m\rangle \sim
p^{a+m}$ as $p \to 0$.

Consider now a generic excited state $\langle\{p_i,q_i\}|=\langle
0|\prod_{i=1}^{N} {\psi}^\dagger_{ q_i } {\psi}_{p_i}$
characterized by a set of $N$-particles with momenta $p_i>a$ and
$N$ holes with momenta $q_i\leq a$. The overlap of this excited
state with the boundary state is obtained from the Wick's theorem and
eq.(\ref{Ga})
\begin{eqnarray}
\langle \{p_i,q_i\}|B_{m}\rangle &=&  \det_{i,j \leq N}\langle 0|
\psi^{ \dag}_{q_{j}} \psi_{p_{i}}  |B_m\rangle =  \det_{i,j \leq
N} \left(f^{+}_m(p_i)\frac{1}{p_i-q_j} f^{-}_m(q_j)\right) \nonumber \\
&=& \prod_{i\leq N}f^{+}_{m}(p_{i})f^{-}_{m}(q_{i}) \;
\frac{\prod_{i<j\leq N} (p_{i}-p_{j})(q_{j}-q_{i})}
{\prod_{i,j\leq N}(p_{i}-q_{j})},  \label{Ga1}
\end{eqnarray}
where in the last line we again used the Cauchy determinant
formula (\ref{C}).

Let us note that  up to the multiplicative factor
$\prod_if^{+}_m(p_i)f^{-}_m(q_i)$,  the overlap (\ref{Ga1})
between the excited state and  the boundary state  written in
momentum space is given by the Cauchy determinant, i.e., has the
same form  as  the  similar overlap written in real space
(\ref{Cauchy1}). The determinants  become identical under a
substitution   $p_i=e^{ix_i}$ and $q_i=e^{iy_i}$.

We observe an interesting phenomenon: the boundary state  $|B_m>$
interchanges momentum and coordinate spaces. The transmutation
\begin{eqnarray}
\mbox{\large  coordinate\; space} &\leftrightarrow&  \mbox{\large
momentum\; space},
\\
\quad \mbox{\large Fermi\; sea} &\leftrightarrow&
\mbox{ \large Boundary \, state}
\end{eqnarray}
is the reason why different matrix elements (\ref{tau1},
\ref{tau2}) obey similar non-linear equations. There is no
one-to-one correspondence between coordinate and momentum space
-- the space is continuous and finite while the momentum space is
discrete and infinite.

Following the logic of the previous section we proceed in two
steps. First, we establish the bosonization formulas in momentum
space. They will allow us to rewrite tau-functions in eq.
(\ref{DH}) as matrix elements of fermionic operators. Then
eq.(\ref{DH}) is equivalent to a particular form of the Wick's
theorem.

\subsubsection*{Bosonization formulas}

Let us introduce a bosonic field associated with the set of commuting Hamiltonians (\ref{HK})
\begin{equation}\label{43}
    \Phi(s)=\sum_{k>0} \frac{s^{-k}}{k}H_k
    = -\sum_{p}\ln(1-\frac{p}{s}):\psi^{\dag}_{p}\psi_{p}:.
\end{equation}

Bosonization in the momentum space is summarized  by formulas
\begin{align}
 \label{BF2}
    {\psi}^\dagger_q {\psi}_p |B_m\rangle
    =\frac{f^{+}_m(p) f^{-}_m(q)}{p-q} e^{\Phi(q)-\Phi(p)} |B_m\rangle,
\end{align}
and
\begin{align}
 \label{BF11}
    & \lim_{p\to -m}\, (p+m){\psi}^\dagger_{q}\psi_p |B_m\rangle
    =(-1)^{m}\frac{\sin\pi a}{\pi}
    f^{-}_{m+1}(q)  e^{ \Phi(q)} |B_{m+1}\rangle.
\end{align}
These formulas will be derived below.  We remark here that
(\ref{BF2},\ref{BF11}) should be understood as analytically
continued formulas for matrix elements. For example, one should
first calculate the overlap of the l.h.s. of (\ref{BF11}) with
some bra-state, then analytically continue it from $p\in {\mathbb
Z}+a$ to the complex $p$-plane and only then take a limit $p\to
-m$.

Once bosonization formulas in momentum space
(\ref{BF2},\ref{BF11}) are established we are ready to prove the
Hirota equation (\ref{DH}).

\subsubsection*{Wick's theorem}
First, we apply the Wick theorem  to the 4-fermion matrix element
\begin{equation}
    \langle{G\bf(t)} | \psi^\dag_q \psi_p \psi^\dag_Q \psi_P | B_m\rangle
    \langle{G\bf(t)}|B_m\rangle
    = \langle{G\bf(t)} | \psi^\dag_q\psi_p | B_m\rangle
    \langle{G\bf(t)} | \psi^\dag_Q \psi_P | B_m\rangle
    - \langle{G\bf(t)} | \psi^\dag_q \psi_P |B_m\rangle
    \langle{G\bf(t)} | \psi^\dag_Q\psi_p | B_m\rangle.
\end{equation}
Next we  multiply both sides by $\frac{P+m}{ f^{-}_{m+1}(Q)} $, send $P\to -m$, and  use eq. (\ref{BF11}). We get
\begin{eqnarray}
    && \frac{q-Q}{p-Q}\langle{G\bf(t)}|e^{ \Phi(Q) }\psi^\dag_q\psi_p|B_{m+1}\rangle
    \langle{G\bf(t)}|B_m\rangle
  \nonumber \\
    &=& \langle{G\bf(t)}|\psi^\dag_q\psi_p|B_m\rangle
    \langle{G\bf(t)}| e^{ \Phi(Q) }|B_{m+1}\rangle
    -\frac{f^{-}_{m+1}(q)}{f^{-}_{m+1}(Q)}
    \langle {G\bf(t)}| e^{\Phi(q)}|B_{m+1}\rangle
    \langle {G\bf(t)}|\psi^\dag_Q\psi_p|B_m\rangle.
\end{eqnarray}
The next step is to use the bosonization formula (\ref{BF2})
\begin{eqnarray}
    & & \frac{q-Q}{p-Q}\; \frac{p+m}{q+m}
    \langle{G\bf(t)}| e^{ \Phi(Q)} e^{\Phi(q)-\Phi(p)} |B_{m+1}\rangle
    \langle{G\bf(t)}|B_m\rangle
  \nonumber \\
    &=& \langle {G\bf(t)} | e^{\Phi(q)-\Phi(p)} | B_m\rangle
    \langle{G\bf(t)}| e^{ \Phi(Q) }|B_{m+1}\rangle
    -\frac{Q+m}{q+m}\; \frac{p-q }{p-Q}
    \langle{G\bf(t)}| e^{ \Phi(q)}|B_{m+1}\rangle
    \langle{G\bf(t)}| e^{\Phi(Q)-\Phi(p)} |B_m\rangle.
 \la{dhi}
\end{eqnarray}
We recognize the matrix elements in (\ref{dhi}) as tau-functions
(\ref{tau2}) with shifted arguments, e.g., $\langle{G\bf(t)}| e^{
\Phi(q)}|B_{m+1}\rangle = \tau_{m+1}(-[q])$. The last step is to
take a limit $Q\to\infty$. In this limit $\Phi(Q)\to 0$ and
(\ref{dhi}) brings us to  (\ref{DH}), which completes the proof.

\subsubsection*{Derivation of formulas for matrix elements (\ref{Ga},\ref{efs})}
Now we prove the formulae (\ref{Ga},\ref{efs}). First, we find
using (\ref{28}) that in coordinate representation
\begin{equation}
    \langle 0|\psi^{\dag}(x)\psi(y)|B_{m}\rangle
    =\langle 0|\psi^{\dag}(x)\psi(y)|0\rangle \left(\frac{1-e^{ix}}{1-e^{iy}}\right)^{m+a}
    =\frac{e^{iy}}{e^{iy}-e^{ix}} \left(\frac{1-e^{ix}}{1-e^{iy}}\right)^{m+a}.
\end{equation}
We consider $(p-q) \langle 0| \psi^{ \dagger }_q \psi_p  |B_{m}\rangle $ and pass to the coordinate representation
$(p-q) \langle 0| \psi^{ \dagger }_q \psi_p  |B_{m}\rangle \to\\ (i\p_x+i\p_y) \langle 0| \psi^{ \dagger  }(x) \psi(y)| B_m\rangle$. Then we calculate
$$
    (i\p_x+i\p_y) \left[\frac{e^{iy}}{e^{iy}-e^{ix}}
    \left(\frac{1-e^{ix}}{1-e^{iy}}\right)^{m+a} \right]
    =(m+a)(1-e^{ix})^{m+a-1} e^{iy}(1-e^{iy})^{-(a+m+1)}.$$
We observe that the right hand size factorizes. Taking an inverse Fourier transform we obtain (\ref{Ga}) with
$$
    f_m^-(q) = \int_0^{2\pi} e^{iqx}(1-e^{ix})^{m+a-1}\,\frac{dx}{2\pi},
    \quad\quad
    f_m^+(p)=(m+a)\int_0^{2\pi} e^{-i(p-1)y}(1-e^{iy})^{-(m+a+1)}\,\frac{dy}{2\pi}.
$$
The calculation of these integrals gives (\ref{efs}) for
$f^\pm_m$. These formulas can also be found in \cite{Schur}.

\subsubsection*{Derivation of bosonization formulas (\ref{BF2},\ref{BF11})}
To prove the formula (\ref{BF2}) we evaluate the projection of its r.h.s.~on the state $\langle\{p_i,q_i\}|$.  With the help of the formula
\begin{equation}
 \label{N}
    e^{b\Phi(s)}{\psi}^\dagger_{q}\psi_p
    e^{-b\Phi(s)}=\left(\frac{s-p}{s-q}\right)^b {\psi}^\dagger_{q}\psi_p
\end{equation}
analogous to (\ref{28})  and using $\langle 0| \Phi(s)=0$ we obtain
\begin{equation}
 \la{rhsBF2}
    \langle\{p_i,q_i\}|e^{\Phi(q)-\Phi(p)}|B_m\rangle
    =\langle\{p_i,q_i\}|B_m\rangle
    \prod_{i\leq N} \frac{ (p-p_i)(q-q_i)}{(p-q_i)(q-p_i)}.
\end{equation}
On the other hand,  using the multiplicative representation for the l.h.s.~of (\ref{BF2}) similar to (\ref{Ga1}) we obtain
\begin{equation}
 \la{lhsBF2}
    \langle\{p_i,q_i\}|{\psi}^\dagger_q {\psi}_p |B_m\rangle
    =\det_{i,j \leq N+1} \left(f^{+}_m(p_i)\frac{1}{p_i-q_j} f^{-}_m(q_j)\right)
    =\langle\{p_i,q_i\}|B_m\rangle
    \frac{f^{+}_m(p) f^{-}_m(q)}{p-q}
    \prod_{i\leq N} \frac{ (p-p_i)(q-q_i)}{(p-q_i)(q-p_i)},
\end{equation}
where $p_{N+1}=p$ and $q_{N+1}=q$. Comparing (\ref{lhsBF2}) and (\ref{rhsBF2}) we prove (\ref{BF2}).

To obtain (\ref{BF11}), we first use (\ref{f}) to obtain
\begin{equation}
 \la{aux}
    \langle\{p_i,q_i\}|B_{m+1}\rangle
    =\langle\{p_i,q_i\}|B_{m}\rangle \prod_{i\leq N}\frac{p_{i}+m}{q_{i}+m}.
\end{equation}
Then we use (\ref{aux}) and (\ref{lhsBF2}) to proceed as follows
\begin{eqnarray}
    \langle\{p_i,q_i\}|e^{\Phi(q)}|B_{m+1}\rangle
    &=&\prod_{i\leq N}\frac{q-q_{i}}{q-p_{i}}\langle\{p_i,q_i\}|B_{m+1}\rangle
    = \prod_{i\leq N}\frac{q-q_{i}}{q-p_{i}}\frac{p_{i}+m}{q_{i}+m}
    \langle\{p_i,q_i\}|B_{m+1}\rangle
 \nonumber \\
    &=& \langle\{p_i,q_i\}|\psi^{\dag}_{q}\psi_{p}|B_{m}\rangle
    \frac{p-q}{f_{m}^{+}(p)f_{m}^{-}(q)}
    \prod_{i\leq N} \frac{p_{i}+m}{p_{i}-p}\frac{q_{i}-p}{q_{i}+m}.
 \la{rhs55}
\end{eqnarray}
The r.h.s. of (\ref{rhs55}) effectively does not depend on $p$ and we take its limit as $p\to -m$.  The product in the last line disappears and we conclude that
\begin{eqnarray}
    e^{\Phi(q)}|B_{m+1}\rangle
    &=&- \frac{q+m}{f_{m}^{-}(q)}
    \lim_{p\to -m}\, \frac{1}{f_{m}^{+}(p)}
    \psi^{\dag}_{q}\psi_{p}|B_{m}\rangle.
\end{eqnarray}
Finally, using
\begin{equation}
    \lim_{p\to -m}\, (p+m)f^{+}_m(p)=(-1)^{m+1}(m+a)\frac{ \sin (\pi a) }{\pi },
\end{equation}
we arrive to (\ref{BF11}).

\section{Multi-periodic solutions}
\la{multiphase}
In this section we present explicit formulas for multi-phase
solutions evolving in real time. These formulas reveal the
structure of the tau-function and provide another direct proof of
Hirota's equation.

A set of excited states $|\{p_i,q_i\} \rangle$ characterized by
momenta of $N$ particles $p_i>a$ and  holes $q_i\leq a,\;i=1,\dots
N$ form an orthonormal basis for the Fock space of fermions. In
this basis the Hamiltonians (\ref{HK}) are diagonal:
$H_k|\{p_i,q_i\}\rangle=\sum_i(p_i^k-q_i^k)|\{p_i,q_i\}\rangle $.
 Using this basis, we decompose
\begin{equation}
\tau_m=\sum_{N\geq 0}\sum_{\{p_i,q_i\}}e^{-\sum_{k,i}
t_k(p_i^k-q_i^k)} \langle {g}|\{p_i,q_i\}\rangle \langle
\{p_i,q_i\}|B_m\rangle.
\end{equation}
The matrix elements entering this expression are determinants of two-particle matrix
elements. The first one is
\begin{equation} \label{Adef}
    \langle{g}|\{p_i,q_i\}\rangle
    =\det_{i,j} A_{p_i,q_j},\quad A_{p,q}=\langle g|p,q\rangle.
\end{equation}
The second one is the Schur function given by eq. (\ref{Ga1})
\begin{equation}
    \langle\{p_i,q_i\}|B_m\rangle
    =\det_{i,j\leq N}\left(\frac{1}{p_i-q_j}\right)
    \prod_{i\leq N} f^{+}_m(p_i) f^{-}_m(q_i).
\end{equation}
Introducing $$s_m(\{p_i,q_i\},g)= \det_{i,j}\left(
A_{p_i,q_j}^{(m)}\right),$$ where $$ A_{p,q}^{(m)}=f^{+}_m(p)
A_{p,q}f^{-}_m(q) $$ we write the tau-function in the form
\begin{equation}\label{MKP3}
\tau_m=\sum_{N\geq 0}\sum_{\{p_i,q_i\}} e^{-\sum_{k,i}
t_k(p_i^k-q_i^k)}s_m(\{p_i,q_i\},g)\det_{ij}
\left(\frac{1}{p_i-q_j}\right).
\end{equation}
It is obvious from this form and inequality $p_i>a\geq q_i$ that
the tau function has only positive Fourier components 
with respect to all odd flows $it_{1},it_{2},\ldots$. Indeed, for all odd $k$ we haver $p_{i}^{k}-q_{i}^{k}>0$. The bilinear
identity (\ref{ContinuousHirotaH}) follows directly from this
representation if one uses  identities for Schur  functions  and
the  property
\begin{equation} \label{sref}
    \frac{s_{m+1}} {s_m} = \prod_i \left( \frac{p_i+m} {q_i+m}
    \right).
\end{equation}
We do not present this proof of (\ref{DH}) here  \footnote{ An
observation that the series (\ref{MKP3}) for (\ref{tau2}) obeys
KP-hierarchy, i.e., eq. (\ref{ContinuousHirotaH})  at $n=0$   has
been made in Ref.\cite{orlov-2003-}. }.

If the matrix $A_{pq}$ defining the state $\langle g|$ by
(\ref{Adef}) is a finite matrix, the  sum in (\ref{MKP3})
truncates. It represents a multi-phase
solution. In this case the determinant formula holds
\begin{equation} \label{dettau}
    \tau_m(\mathbf{t},a)=\det_{i,j\leq N}\left(\delta_{ij}+f^{+}_m(p_i) e^{-\epsilon
    (p_i)}\frac{A_i}{p_i-q_j}e^{\epsilon (q_j)}f^{-}_m(q_j)\right),
    \quad \epsilon(p)=\sum_k t_k p^k.
\end{equation}
Here $A_{i}=A_{p_{i}q_{i}}$. Note that a similar determinant representation holds for
(\ref{tau1}). Namely,
\begin{equation} \label{dettau1}
    \tau_m(\mathbf{t})=\det_{i,j\leq N}\left(\delta_{ij}+ p_{i}^{-m}
    e^{-\epsilon(p_i)}\frac{A_i}{p_i-q_j}e^{\epsilon (q_j)}q_{j}^{m+1}\right),
    \quad \epsilon(p)=\sum_k t_k p^k.
\end{equation}
Although (\ref{dettau}) and (\ref{dettau1}) are very similar there is an essential difference between them. Namely, in (\ref{dettau}) the parameters $p_{i}$ and $q_{i}$ are momenta of particles and holes and are subject to ``Fermi sea'' restriction $p_{i}>a\geq q_{i}$. On the other hand, in (\ref{dettau1}) $p=e^{ix}$ and $q=e^{iy}$ represent complex
coordinates of particles and holes and are not subject to this restriction. 
Due to the restriction on the values of $p_{i}$ and $q_{i}$ the tau function (\ref{dettau}) has only positive Fourier components in all odd flows. 

The simplest 1-phase and 2-phase solutions are written below
as a function of space $t_{1}=ix$ and time $t_{2}=-it$.
\begin{itemize}
\item[-] 1-phase solution describes one particle-hole
excitation:
\begin{equation} \label{1}
\tau_m=1+A \frac{f^{+}_m(p) f^{-}_m(q)} {p - q} e^{i(p-q)(x-vt)}.
\end{equation}
The solution is characterized by two positive numbers $p$ and $-q$
from ${\mathbb Z}+a$ labelling particle-hole momenta and an
arbitrary $m$-independent constant $A$, which may depend on $p,q$.
It propagates with velocity $v=p+q$.
\item[-] 2-phase
solution (two particle-hole excitations):
\begin{align} \label{2}
& \tau_m = 1 + A_1 \frac{ f^{+}_m(p_1) f^{-}_m(q_1) } { p_1 - q_1}
e^{ i(p_1-q_1)x - i(p_1^2-q_1^2)t} + A_2 \frac{ f^{+}_m(p_2)
f^{-}_m(q_2) } { p_2 - q_2} e^{i (p_2 - q_2)x - i( p_2^2 - q_2^2)
t} \\
& +  A_1 A_2\det_{i,j=1,2}
\left(\frac{f^{-}_m(q_j)f^{+}_m(p_i)}{p_i-q_j} \right) e^{i(p_1  -
q_1) (x - v_1 t)}  e^{i( p_2 - q_2)( x - v_2 t)}. \nonumber
\end{align}
\end{itemize}

The multi-phase solutions in (\ref{dettau1}) allow for rational
limit. Keeping $v_{i}=p_{i}+q_{i}$ constant while taking the limit
$p_{i}-q_{i}\to 0$, one obtains rational solutions corresponding
to $N$-solitons with $v_{i}$ being asymptotic velocities of
solitons. However, in (\ref{dettau}) and its particular cases
$(\ref{1},\ref{2})$ this limit does not exist. Indeed, under the
condition $p_{i}>a\geq q_{i}$ the limit $p_{i}-q_{i}\to 0$ implies
$p_{i},q_{i}\to 0$ and therefore it is impossible to have finite
$v_{i}$ in this limit. We see that the matrix element (\ref{tau2})
does not contain soliton solutions of the MKP hierarchy.

\section{Conclusion}

We have shown that the real time dynamics  of certain matrix
elements (\ref{tau2}) of  one-dimensional (non-relativistic)
fermions obeys non-linear integrable equations of the modified KP
hierarchy. This evolution is driven by a commutative set of
Hermitian Hamiltonians  (\ref{HK}) $$H_k=\int
:\psi^\dag(x) \left(-i\frac{\p}{\p x} \right)^k \psi(x)
:\frac{dx}{2\pi}. $$

It has  to be contrasted with a traditional representation of the
tau-function of the modified KP - hierarchy (\ref{tau1})
\cite{1,2} in terms of fermions. The latter describes an evolution
driven by a commutative set of (non-Hermitian) current operators
(\ref{J}) $$J_k=\int e^{ikx} :\psi^\dag(x) \psi(x)
:\frac{dx}{2\pi}.$$ In these two cases the flows have different
meaning. The former are coordinates of physical space-time, while
the latter are deformation parameters of the coherent state.

The matrix element  we studied in this paper is understood as an
overlap of the evolved coherent fermionic state $\langle
G(\mathbf{t})|$ with a ``probe'' boundary state $|B_{m}\rangle$.
This matrix element appears in variety of physical problems
involving the phenomena referred to as the Orthogonality
Catastrophe. Among them are problems related with the Fermi-edge
singularity, quantum impurity and tunneling in  effectively one
dimensional systems.

The fact that questions of quantum dynamics in one dimension are
linked to the tau-function, indicates that the kinetics are
essentially non-linear and are a subject to  instabilities
inherited from the non-linear dynamics. In a forthcoming paper
\cite{BAW} we describe the Whitham approach to shock wave type
solutions for electronic systems.

An essential element of our  proof of the MKP hierarchy is the
representation of a fermionic mode  by a Bose field
(\ref{43},\ref{BF11})  in the action on a boundary state,
$$\psi_p \sim f^+(p)e^{ -\Phi(p)}.$$ This is a representation
in momentum space. It is  valid in a subspace of the Fock space
consisting of excitations on  top of the boundary state.

Again it has to be contrasted with the traditional ``bosonization"
- representation of Fermi-field  in terms of a Bose-field with
respect to the Fermi-vacuum $$\psi(x)\sim e^{\varphi(x)}.$$ This
is a representation in coordinate space. It is  valid in a
subspace of the  Fock space consisting of excitations on  top of
the Fermi-vacuum.

We emphasize that although the matrix elements (\ref{tau1}) and
(\ref{tau2}) obey similar MKP equations they give essentially
different solutions of those equations. For example, (\ref{tau2})
gives only solutions satisfying analyticity requirements in odd
flows. The latter condition excludes soliton solutions.

Finally, we  emphasize  that the Schur decomposition (\ref{MKP3})
of the tau-functions similar to  (\ref{tau2}) (an  essential
element of the proof) has appeared in \cite{okounkov-2000-} in the
problem of counting of Hurwitz numbers and in \cite{orlov-2003-}
in constructions of tau-functions of hypergeometric type. Similar
objects arise in  supersymmetric gauge theories \cite{N,okounkov-2000-} and also
in matrix quantum mechanics.

\section{Acknowledgement}

We have  benefited from discussions with   I. Krichever,  A.
Orlov,  I. Gruzberg, J. Harnad, T. Takebe, A. Zabrodin. We thank
A. Zabrodin and J. Harnard who pointed out the paper
\cite{orlov-2003-} to us and N. Nekrasov for the information about
\cite{okounkov-2000-}. P.W.  and E.B. were supported by the NSF
MRSEC Program under DMR-0213745 and NSF DMR-0220198. E.B. was also
supported by BSF  2004128 P.W. acknowledges support  by the
Humboldt foundation and  is grateful to Takashi Takebe for his
kind hospitality in Ochanomizu University. The work of AGA was
supported by the NSF under the grant DMR-0348358.

\appendix
\section{Reduction from the particle-hole symmetry}
\la{app:reduction}

Let us treat  the tau-function as a function of space-time
(\ref{t}). At first glance, the equation (\ref{MKP1})  is not
closed. It is written on two functions $\tau_m$ and $\tau_{m+1}$.
In fact, both functions are determined by the initial state
$\langle g|$ and are not independent. The relation between these
functions determined by formulas (\ref{MKP3},\ref{sref}) makes
(\ref{MKP1}) a closed equation.

However, in the case of generic $g$ the relations between $\tau_m$
and $\tau_{m+1}$ are difficult to formulate. The situation is
simplified when the initial state $\langle g|$ has some special
form, or  is restricted by symmetries. One may think about these
cases as \textit{reductions} of MKP hierarchy.

The multi-phase solution (\ref{MKP3})  is the simplest (finite-dimensional)
reduction. In this case the tau-function, treated as a  function
of the variable $z=e^{-ix}$ is a  polynomial in $z^{-1}$.

A more interesting reduction having direct physical applications
occurs when the parameter $a+m$  takes a ``symmetric'' value such
that $a+m=-(a+m+1)$. We set\footnote{This case  is
known as a ``unitary limit'' in problems related to Fermi-edge
singularities.} $$a=-1/2,\quad m=0.$$ In addition we require  the
state $\langle g|$ to be  invariant under a charge  conjugation
\begin{equation} \label{BO1}
e^{2ix_0(p-q)} \langle g|p,q \rangle = - \overline{\langle g| -q,
-p \rangle},
\end{equation}
where $x_0$ is some real constant. This condition reflects a a
particle-hole symmetry:  $\psi_{q}^{\dag} \to \psi_{-q}$, $\psi_{
p } \to \psi_{ -p }^{ \dag }$.

Under these conditions the tau-function treated  as a function of
real space-time coordinates (\ref{t}) has a reflection property
\begin{equation} \label{S}
\tau_0(x)=\overline{\tau_{1}(-x+2x_0)},
\end{equation}
where $\overline{\tau}$ stands for complex conjugation of the
function.  This property makes the equation (\ref{MKP1}) closed
with the second complex function $\tau_{1}$ being completely
determined by $\tau_{0}$.

The analytical meaning of the  condition (\ref{S}) becomes clear
if one considers an extension of the tau-function to the complex
plane $z$ from the unit  circle $z=e^{-i(x-x_0)}$. Then the
reflection condition  means that  the coefficients of the formal
series  (\ref{MKP3}) for $\tau_0$ and $\tau_{1}$ are complex
conjugated  (we use a notation $\overline{f} (z) = \overline{ f(
\bar{z} ) } $)
\begin{equation} \label{S1}
\tau_0(z)=\overline{\tau_{1}}(z).
\end{equation}
This follows from the property
$$ f_{1}^+ ( p )f_{1}^-( q ) = f_{0}^+(-q) f_{0}^-(-p)$$ held at $a=-1/2$.

The holomorphic  tau-function $\tau(z)=\tau_0(z)$ is defined in
the exterior of the unit disk, while its Schwarz reflection
$\tau(\bar z) = \bar\tau_{1}(1/\bar z)$ s defined in the interior
of the disk. The reflection condition (\ref{S1}) means that the
holomorphic tau-function, and its Schwarz reflection, can be glued
along the unit circle forming a function  analytical in some
neighborhood of the circle.

Particle-hole symmetry (\ref{BO1}) occurs when $A_{pq}$ is a
Toeplitz matrix, i.e., it depends on the difference $p-q$. In this
case the state $\langle g|$ is given by
\begin{equation}\label{68}
g=e^{i\int A(x) \rho(x) dx },\quad A(x-x_0)=-A(-x+x_0),
\end{equation}
where $A(x)$ - the symbol of the Toeplitz matrix i.e., a Fourier
transform of $A_{pq}$. It is an odd real function.




\begin{thebibliography}{10}



\bibitem{1}

Etsuro Date, Michio Jimbo, Masaki Kashiwara, and Tetsuji Miwa.
\newblock Transformation groups for soliton equations.
\newblock In M.~Stone, editor, {\em Bosonization}, Singapore, 1994. World Scientific.
\newblock RIMS-394.



\bibitem{Sato1}
M.~Sato.
\newblock Soliton equations as dynamical systems on infinite dimensional
{G}rassmann manifolds.
\newblock {\em RIMS Kokyuroku}, 439:30--40, 1981.
\newblock Sato, Mikio and Sato, Yasuko: Soliton equations as dynamical
systems on infinite-dimensional Grassmann manifold. Nonlinear
partial differential equations in applied science (Tokyo, 1982),
259--271, North-Holland Math. Stud., 81, North-Holland, Amsterdam,
1983.


\bibitem{2}
E.~Date T.~Miwa, M.~Jimbo.
\newblock {\em {S}olitons: Differential Equations, Symmetries and Infinite
Dimensional Algebras}, volume 135 of {\em Cambridge Tracts in Math.}
\newblock Cambridge university press, London, Great Britain, 1999.
\newblock Michio Jimbo and Tetsuji Miwa.
\newblock {S}olitons and infinite dimensional lie algebras.
\newblock {\em Publ. Res. Inst. Math. Sci. Kyoto}, 19:943, 1983.

E.~Date, M.~Jimbo, and T.~Miwa.
\newblock Method for generating discrete {S}oliton equations. {I}.
\newblock {\em Journal of the Physical Society of Japan}, 51 (12):4116--4124,
1982.

E.~Date, M.~Jimbo, and T.~Miwa.
\newblock Method for generating discrete {S}oliton equations. {II}.
\newblock {\em Journal of the Physical Society of Japan}, 51 (12):4125--4131,
1982.

E.~Date, M.~Jimbo, and T.~Miwa.
\newblock Method for generating discrete {S}oliton equations. {III}.
\newblock {\em Journal of the Physical Society of Japan}, 52 (2):388--393,
 1983.






\bibitem{SIK}
A.~R. {Its}, A.~G. {Izergin}, V.~E. {Korepin}, and N.~A. {Slavnov}.
\newblock {Differential Equations for Quantum Correlation Functions}.
\newblock {\em International Journal of Modern Physics B}, 4:1003--1037, 1990.


\bibitem{SMJ}
M.~{Jimbo}, T.~{Miwa}, Y.~{M{\^o}ri}, and M.~{Sato}.
\newblock {Density matrix of an impenetrable Bose gas and the fifth
 Painlev{\'e} transcendent}.
\newblock {\em Physica D Nonlinear Phenomena}, 1:80--158, April 1980. M. Sato, T. Miwa, and M. Jimbo, Proc. Jpn. Acad. 53A, 147, 153, 
183 (1977); Publ. Res. Inst. Math. Sci. 14, 223 (1978); 15, 201, 577, 871 (1979).
\bibitem{orlov-2003-}

A~Yu Orlov.
\newblock Hypergeometric tau functions $\tau({\bf t},t,{\bf t}^*)$ as
$\infty$-{S}oliton tau function in t variables.
\newblock arXiv.org:nlin/0305001.



\bibitem{okounkov-2000-}
A. Okounkov.
\newblock Toda equations for Hurwitz numbers.
\newblock arXiv:math/0004128.
A. Okounkov and R. Pandharipande.
\newblock The equivariant Gromov-Witten theory on $\mathbf{P}^1$.
\newblock arXiv:math.AG/0207233.
N. A. Nekrasov and A. Okounkov.
\newblock Seiberg-Witten theory and random partitions.
\newblock arXiv:hep-th/0306238.

\bibitem{AA}
I.~{Affleck} and A.~W.~W. {Ludwig}.
\newblock {The Fermi edge singularity and boundary condition changing operators.}
\newblock {\em Journal of Physics A Mathematical General}, 27:5375--5392,
 August 1994.
\bibitem{Ueno-Takasaki}K.Ueno and K.Takasaki, Toda lattice hierarchy, Adv. Studies in Pure Math. 4 (1984) 1-95.

\bibitem{OC} P. Nozires and C. T. deDominicis,
Phys. Rev. 178, 1097 (1969).
G. D. Mahan, Phys. Rev. 163, 612 (1967) .
P. W. Anderson, Phys. Rev. Lett. 18, 1049 (1967) .


\bibitem{Schur}
I.G.Macdonald, 
\newblock Symmetric Functions and Hall Polynomials, 
\newblock Clarendon Press, 
\newblock Oxford, 1995

\bibitem{Hirota:PRL}
R.~Hirota.
\newblock Exact solution of the korteweg—de vries equation for multiple
collisions of solitons.
\newblock {\em Phys. Rev. Lett.}, 27:1192, 1971.

\bibitem{Satsuma}
J.~Satsuma and Y.~Ishimori.
\newblock Periodic-wave and rational {S}oliton solutions of the
{B}enjamin-{O}no equation.
\newblock {\em J. Phys. Soc. Japan}, 46:681--687, 1979.

\bibitem{BAW} E. Bettelheim, A. G. Abanov and P. Wiegmann, Quantum Schock Waves - the case for non-linear effects in dynamics of electronic liquid.
\newblock arXiv:cond-mat/0606778;
\newblock Orthogonality catastrophe and shock waves in a non-equilibrium Fermi gas.
\newblock arXiv:cond-mat/0607453.

\bibitem{N} Nekrasov  N.A.: Seiberg-Witten Prepotential from Instanton Counting, hep-th/0206161 v1;
Losev A, S., Marshakov A. and Nekrasov N.A.: Small Instantons, Little 
Strings and Free Fermions, hep-th/0302191 v3. 

\end{thebibliography}
\end{document}